\documentclass[nofootinbib]{revtex4}
\usepackage{graphicx}
\usepackage{fancyhdr}
\usepackage{amsmath,amssymb}
\usepackage{color}

\begin{document}

\title{
$\tan\beta$ determination from the Higgs boson decay at the
International Linear Collider\footnote{The talk is based on
Ref.~\cite{Kanemura:2013eja}.} }


\author{Hiroshi Yokoya}
\affiliation{Department of Physics, University of Toyama, Toyama
930-8555, JAPAN}

\begin{abstract}
We study the methods and their accuracies for determining $\tan\beta$ in
 two Higgs doublet models at future lepton colliders.
In addition to the previously proposed methods using direct
 production of additional Higgs bosons, we propose a method using the
 precision measurement of the decay branching ratio of the
 standard-model (SM)-like Higgs boson.
The method is available if there is a deviation from the SM in the
 coupling constants of the Higgs boson with the weak gauge bosons.
We find that, depending on the type of Yukawa interactions, this method
 can give the best sensitivity in a wide range of $\tan\beta$.
\end{abstract}

\maketitle

\thispagestyle{fancy}


%%%%%%%%%%%%%%%%%%%%%%%%%%%%%%%%%%
\section{Introduction}

A Higgs boson has been discovered at the LHC~\cite{Ref:atlas,Ref:cms}.
No indication has been found so far for the deviation from the standard
model~(SM) on the nature of the Higgs boson such as the decay width, 
spin-parity, and also on the coupling constants with SM fermions up to
current experimental accuracies~\cite{Aad:2013wqa,Aad:2013xqa,%
Chatrchyan:2013iaa,Chatrchyan:2013mxa}. 
In spite of such situation, Higgs sector is not yet settled to be
composed with one Higgs doublet. 
Namely, there are various other possibilities which are still consistent
with current experimental data, and even more, plenty of models with the
extended Higgs sector have been proposed to explain the phenomena which
may indicate physics with a new energy scale beyond the SM, such as
hierarchy problem, neutrino masses, dark matter, etc. 

Searches for the evidence of extended Higgs sectors are of primary
importance at future experiments. 
At the second stage of the LHC experiment with $\sqrt{s}=13$ or 14~TeV,
direct searches for the evidence of additional Higgs bosons can be
performed and the energy reach for new particles will be extended. 
On the other hand, precise measurements of the couplings of the Higgs
boson can be performed at the future International Linear Collider (ILC)
experiment~\cite{Asner:2013psa,Dawson:2013bba}, and the evidence of
non-standard Higgs models can be detected as a deviation from the SM in
the coupling constants of the SM-like Higgs boson. 
Furthermore, some parameter regions where the direct searches at the LHC
cannot be covered can be complemented by the searches at the
ILC~\cite{Kanemura:2014dea}.
The model discrimination can be performed through the direct measurement
of the properties of additional Higgs bosons and/or fingerprinting the
pattern of the deviation in various coupling measurements~\cite{KTYY}. 

In this talk, we discuss collider methods for the determination of
$\tan\beta$, a ratio of vacuum expectation values of the two doublets,
in the two Higgs doublet model (THDM) as a benchmark model for the
extended Higgs sector. 
We propose a new method through the measurements of the branching ratio
of the SM-like Higgs boson at future lepton colliders.
The method is applicable as long as there exist deviations in the
couplings of the SM-like Higgs boson to gauge bosons, even for the case
where additional Higgs bosons are too heavy to be detected directly.
We studies the sensitivity of determining $\tan\beta$ at the ILC, and
compare it with those for the previously proposed methods which utilize
direct production of additional Higgs bosons~\cite{Ref:TanB}. 

%%%%%%%%%%%%%%%%%%%%%%%%%%%%%%%%%%
\section{Two Higgs Doublet Model}

In this section, we briefly review the THDM with a softly-broken
discrete $Z_2$ symmetry.
This model has two preferable features which are good to naturally avoid
phenomenological constraints on the extended Higgs sector. 
One is that any multi-doublet model predicts the electroweak rho
parameter, $\rho=m_W^2/(m_Z^2\cos^2\theta_W)$, to be unity at the tree
level. 
Since the experimental constraint on the rho parameter is quite strict, 
$\rho_{\rm exp}=1.0004^{+0.0003}_{-0.0004}$~\cite{Beringer:1900zz},
these models may be regarded as natural extension of the SM.
Second is that the $Z_2$ symmetry can suppress the flavor changing
neutral currents (FCNCs) which are also severely constrained by flavor
experiments. 
Under the $Z_2$ symmetry, each fermion couples to only one Higgs
field, so that the Higgs-mediated FCNCs are prevented at the tree
level and the constraints are relaxed to the loop level~\cite{Ref:GW}.

In the THDM with $Z_2$ symmetry, depending on the assignment of the
$Z_2$ parity to each fermion, four types of Yukawa interaction can be
constructed~\cite{Barger:1989fj,Grossman:1994jb,Ref:AKTY}. 
Among the four types, we focus on the so-called Type-II and
Type-X~(lepton specific) THDMs, since these deserve much interests from
the viewpoint of constructing models for physics beyond the SM.
Type-II THDM is well-known as the Higgs sector in the minimal
supersymmetric extension of the SM, where up-type quarks couple to one
Higgs doublet while down-type quarks and charged leptons couple to
another Higgs doublet.
Type-X THDM is sometimes employed in models for neutrino masses, etc.,
where quarks couple to one Higgs doublet while charged leptons couple to
another Higgs doublet.
%Similar discussion on the other two types, so-called Type-I and
%Type-Y~(flipped) THDMs can be found in Ref.~\cite{Kanemura:2013eja}.

For simplicity, we restrict ourselves to consider the CP-conserving
scenario, where CP-even $H$, CP-odd $A$ and charged $H^\pm$ Higgs bosons
appear as mass eigenstates in addition to the light CP-even $h$ which we
assume the observed Higgs boson with $m_h=125$~GeV.
Mixing angles $\alpha$ and $\beta$ are defined, respectively, as the
angles in the neutral CP-even states and that between $A$ and $z$ the
neutral component of Nambu-Goldstone boson, as well as between $H^\pm$
and $w^\pm$ the charged components of Nambu-Goldstone bosons.
$\beta$ satisfies $\tan\beta=v_2/v_1$ where $v_i$ are vacuum expectation
values of the two Higgs fields.

\begin{table}[t]
\begin{center}
\begin{tabular}{|c||c|c|c|c|c|c|c|c|c|}
\hline
& $\xi_h^u$ & $\xi_h^d$ & $\xi_h^\ell$
& $\xi_H^u$ & $\xi_H^d$ & $\xi_H^\ell$
& $\xi_A^u$ & $\xi_A^d$ & $\xi_A^\ell$ \\ \hline
Type-I
& $c_\alpha/s_\beta$ & $c_\alpha/s_\beta$ & $c_\alpha/s_\beta$
& $s_\alpha/s_\beta$ & $s_\alpha/s_\beta$ & $s_\alpha/s_\beta$
& $\cot\beta$ & $-\cot\beta$ & $-\cot\beta$ \\
Type-II
& $c_\alpha/s_\beta$ & $-s_\alpha/c_\beta$ & $-s_\alpha/c_\beta$
& $s_\alpha/s_\beta$ & $c_\alpha/c_\beta$ & $c_\alpha/c_\beta$
& $\cot\beta$ & $\tan\beta$ & $\tan\beta$ \\
Type-X
& $c_\alpha/s_\beta$ & $c_\alpha/s_\beta$ & $-s_\alpha/c_\beta$
& $s_\alpha/s_\beta$ & $s_\alpha/s_\beta$ & $c_\alpha/c_\beta$
& $\cot\beta$ & $-\cot\beta$ & $\tan\beta$ \\
Type-Y
& $c_\alpha/s_\beta$ & $-s_\alpha/c_\beta$ & $c_\alpha/s_\beta$
& $s_\alpha/s_\beta$ & $c_\alpha/c_\beta$ & $s_\alpha/s_\beta$
& $\cot\beta$ & $\tan\beta$ & $-\cot\beta$ \\
\hline
\end{tabular}
\end{center}
\caption{The scaling factors for the four types of Yukawa
 interactions in the THDM~\cite{Ref:AKTY}.} \label{Tab:sf}
\end{table}

Coupling constants of $\Phi VV$ interactions, where $\Phi=H$ or $h$, are
given as 
\begin{align}
g^{\rm THDM}_{hVV}=g^{\rm SM}_{hVV}\cdot\sin(\beta-\alpha),\quad
g^{\rm THDM}_{HVV}=g^{\rm SM}_{hVV}\cdot\cos(\beta-\alpha),
\end{align}
where $g^{\rm SM}_{hVV}$ is the corresponding coupling constant for the
SM Higgs boson.
When $\sin(\beta-\alpha)=1$, which is called ``SM-like
limit''~\cite{Gunion:2002zf}, $h$ has the same coupling constants with
gauge bosons as those of the SM Higgs boson.
In general, $\sin(\beta-\alpha)$ is a free parameter in the model.
However, a large deviation of $\sin(\beta-\alpha)$ from unity is
restricted by theoretical constraints which are derived by using the
argument of perturbative unitarity~\cite{Ref:Uni-2hdm}.
Experimental constraints have also been obtained at the
LHC~\cite{Aad:2013wqa,CMS:yva,ATLAS:2013zla,CMS:2013eua}.

The Yukawa couplings for each type in the THDM are characterized by a
scaling factor, $\xi_\phi^f$, defined as the coupling constant in each
model divided by that for the coupling constant in the SM.
For example, 
\begin{align}
 \xi_h^f=\sin(\beta-\alpha)+\cot\beta\cdot\cos(\beta-\alpha),\\
 \xi_H^f=\cos(\beta-\alpha)-\cot\beta\cdot\sin(\beta-\alpha),
\end{align}
for $f=u$ in Type-II and $f=u,d$ in Type-X, while
\begin{align}
 \xi_h^f=\sin(\beta-\alpha)-\tan\beta\cdot\cos(\beta-\alpha),\\
 \xi_H^f=\cos(\beta-\alpha)+\tan\beta\cdot\sin(\beta-\alpha),
\end{align}
for $f=d,\ell$ in Type-II and $f=\ell$ in Type-X.
Thus, in the SM-like limit, Yukawa couplings for $h$ become the same as
those in the SM, and their $\tan\beta$ dependence disappears. 
On the other hand, $\tan\beta$ dependence on the Yukawa couplings for
$H$, as well as $A$ and $H^\pm$, remains in this limit.
The scaling factors are summarized in Table.~\ref{Tab:sf} for the four
types of Yukawa interaction in the THDM.

\begin{figure}[t]
 \centering
 \includegraphics[width=0.25\textwidth]{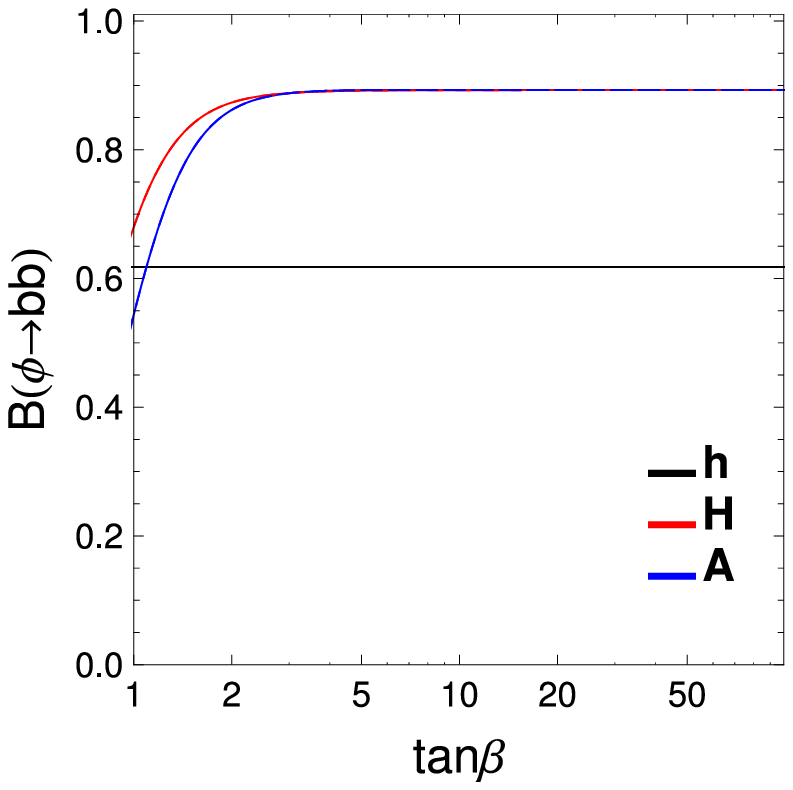} 
 \includegraphics[width=0.25\textwidth]{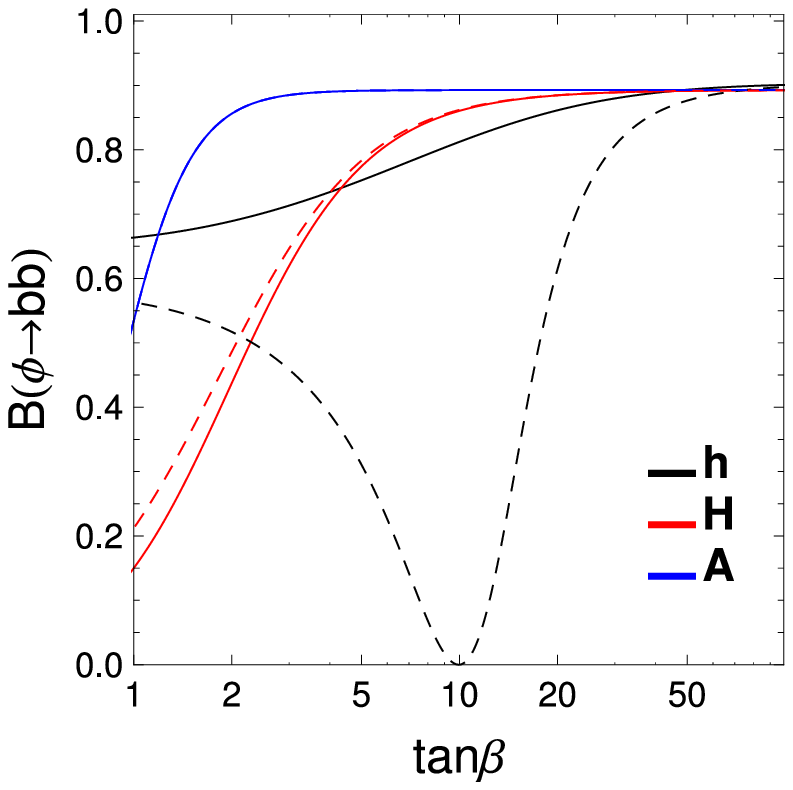} 
 \includegraphics[width=0.24\textwidth]{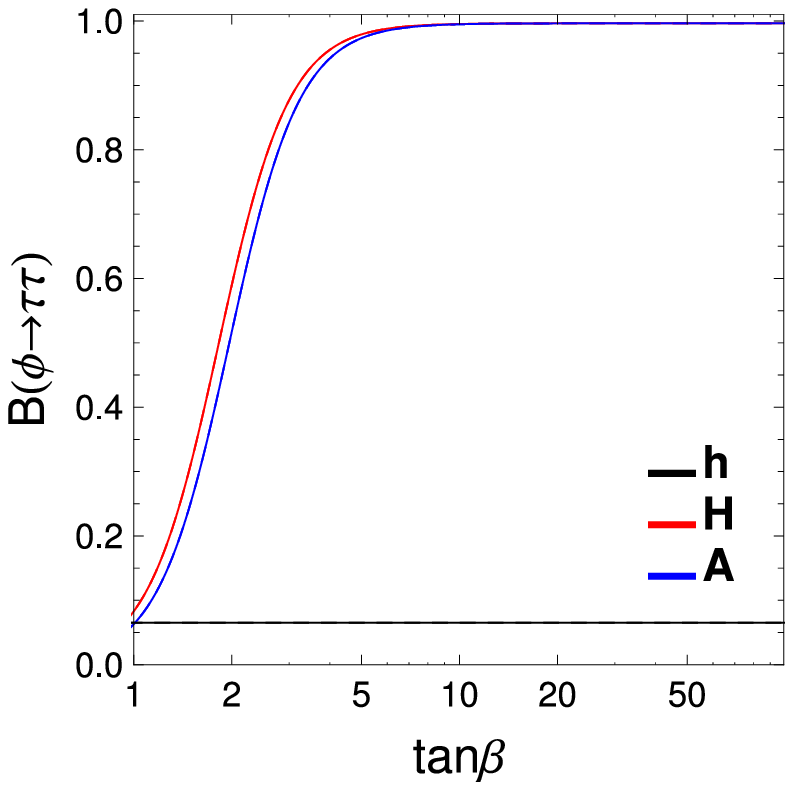} 
 \includegraphics[width=0.24\textwidth]{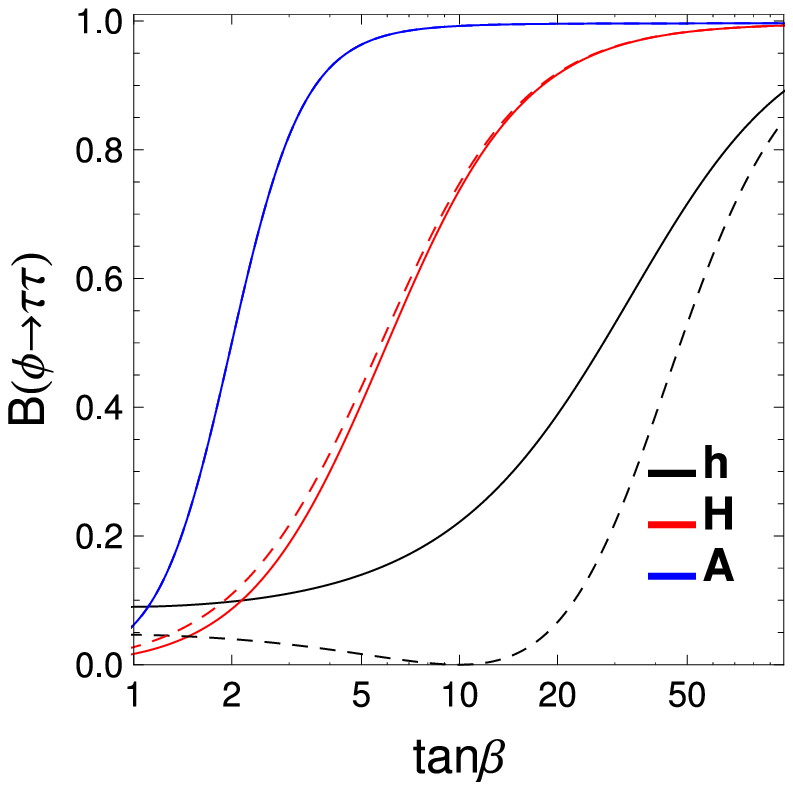} 
 \caption{Left two panels: the decay branching ratios for $h\to b\bar b$
 (black curves), $H\to b\bar b$ (red curves), and $A\to
 b\bar b$ (blue curves) decays as a function of $\tan\beta$ in the
 Type-II THDM with $\sin^2(\beta-\alpha)=1$ and 0.99, respectively.
 The solid (dashed) curves denote the case with $\cos(\beta-\alpha) \le 0$
 ($\cos(\beta-\alpha) \ge 0$). 
 Right two panels: the same as left two panels, but for the
 $\tau^+\tau^-$ decays in the Type-X THDM. }
 \label{Fig:bb}
\end{figure}

In Fig.~\ref{Fig:bb}, we show branching ratios of neutral Higgs bosons
in the $b\bar{b}$ and $\tau^+\tau^-$ decay modes for $m_h=125$~GeV and
$m_H=m_A=200$~GeV in the Type-II and Type-X THDM.
From the left, ${\mathcal B}(b\bar{b})$ for Type-II with
$\sin^2(\beta-\alpha)=1$, that with $\sin^2(\beta-\alpha)=0.99$, 
${\mathcal B}(\tau^+\tau^-)$ for Type-X with
$\sin^2(\beta-\alpha)=1$, that with $\sin^2(\beta-\alpha)=0.99$ are
plotted as a function of $\tan\beta$, respectively.
Solid~(dashed) lines are for
$\cos(\beta-\alpha)<0$~($\cos(\beta-\alpha)>0$).
We see that, when $\sin^2(\beta-\alpha)=1$, branching ratios of $h$ are
independent of $\tan\beta$. 
However, once it deviates from unity, there appear substantial
$\tan\beta$ dependence, and the branching ratios vary in a wide range. 
$\tan\beta$ dependence on the branching ratios of $H$ and $A$ is also
large, and it remains even in the SM-like limit.
%

%%%%%%%%%%%%%%%%%%%%%%%%%%%%%%%%%%
\section{$\tan\beta$ measurement}

In this section, we discuss methods of $\tan\beta$ determination in
the future lepton colliders. 
We consider three methods which utilize the measurements of the
following observables, respectively,\footnote{%
The other method by using the cross-section measurements of
$b\bar{b}H+b\bar{b}A$ production is also proposed in
Ref.~\cite{Ref:TanB}.
This method may become useful for heavier $H$ and $A$ case where $H$ and
$A$ cannot be produced in pair but only singly due to the kinematical
limitation.}
(i) branching ratios of $H$ and $A$, ${\mathcal B}_{H,A}$,
(ii) total decay widths of $H$ and $A$, $\Gamma_{H,A}$,
(iii) branching ratio of $h$, ${\mathcal B}_h$.
The first two observables can be studied in the direct production of $H$
and $A$, i.e., the $e^+e^-\to HA$ process.
Thus, these can be available if the sum of the mass of $H$ and $A$
is less than the collider energy.
Since the production cross section is independent of model parameters,
and the branching ratio of $H$ and $A$ in the $b\bar{b}$ and
$\tau^+\tau^-$ decay modes significantly depend on $\tan\beta$,
$\tan\beta$ can be determined by counting the number of events for
$4b$~($4\tau$) events; ${\mathcal N}\propto\sigma_{HA}\cdot{\mathcal
B}_H\cdot{\mathcal B}_A$. 
Thus, the observation of the branching ratios gives $\tan\beta$
determination by comparing with the theoretical prediction.
We note that masses of $H$ and $A$ can be easily measured from the peak
in the invariant mass distribution.

The last method utilizes the precision measurement of the branching
ratio of $h$.
In the THDM, as we see in Eqs.~(2-5), when $\sin(\beta-\alpha)<1$, the
Yukawa couplings of $h$ can be deviated from those in the SM.
It is known that the pattern of the deviations for up-type, down-type
quarks and charged leptons depends on the type of Yukawa interaction,
therefore, by observing it we could distinguish the type of Yukawa
interaction in the THDM~\cite{KTYY,Kanemura:2014dja}. 
Furthermore, the magnitude of the deviation depends on the value of
$\tan\beta$, so that we can determine $\tan\beta$ by observing it.
The accuracy of the $\tan\beta$ determination depends on how accuracy the
branching ratio can be measured experimentally and also how steeply the
branching ratio depends on $\tan\beta$.

%%%%%%%%%%%%%%%%%
\section{Results}

In this section, we study the accuracies of $\tan\beta$ measurement
for the above three methods at the ILC.

For the method~(i), the sensitivity is estimated as follows.
We utilize $b\bar{b}$ decay mode for Type-II and $\tau^+\tau^-$ decay
mode for Type-X which are basically large and also which have large
$\tan\beta$ dependence.
The expected number of events for $4b$ and $4\tau$ events can be
obtained as $N=\sigma_{HA}\cdot{\mathcal B}_H\cdot{\mathcal
B}_A\cdot{\mathcal L}\cdot{\mathcal \epsilon}$, where
$\epsilon$ is the acceptance ratio for observing $4b$ and $4\tau$
signals. 
We take $m_H=m_A=200$~GeV and $\sqrt{s}=500$~GeV with ${\mathcal 
L}=250$~fb$^{-1}$. 
$\epsilon_{4b}$ and $\epsilon_{4\tau}$ are estimated to be 50\% for both
by our simulation~\cite{Kanemura:2013eja}.
The $1\sigma$ sensitivity to $\tan\beta$ is obtained by solving
$N(\tan\beta\pm\Delta\tan\beta)=N_{\rm obs}\pm\Delta N_{\rm obs}$, where
$\Delta N_{\rm obs}=\sqrt{N_{\rm obs}}$ is a statistical error. 

For the method~(ii), $\tan\beta$ sensitivity from the width measurement
is estimated as follows.
The detector resolutions for the Breit-Wigner width in the $b\bar{b}$
and $\tau^+\tau^-$ invariant mass distributions are estimated to be
$\Gamma^{\rm res}_{b\bar{b}}=11$~GeV and $\Gamma^{\rm
res}_{\tau^+\tau^-}=7$~GeV, respectively~\cite{Kanemura:2013eja}. 
The width to be observed is 
\begin{align}
\Gamma^R_{H/A}=\frac{1}{2}\left[\sqrt{\left(\Gamma_{H}^{\rm
 tot}\right)^2+\left(\Gamma^{\rm res}\right)^2}+
\sqrt{\left(\Gamma_{A}^{\rm tot}\right)^2+\left(\Gamma^{\rm
 res}\right)^2}\right], 
\end{align}
and the $1\sigma$ uncertainty is given by~\cite{Ref:TanB}
\begin{align}
\Delta\Gamma^R_{H/A}=\sqrt{\left(\Gamma^R_{H/A}/\sqrt{2N_{\rm
obs}}\right)^2+\left(\Delta\Gamma^{\rm res}_{\rm
sys}\right)^2},
\end{align}
where $\Delta\Gamma^{\rm res}_{\rm sys}$ is taken as 10\% of
$\Gamma^{\rm res}$ for each decay mode. 
Then, $1\sigma$ sensitivity for the $\tan\beta$ determination is 
obtained by solving $\Gamma_{H/A}(\tan\beta\pm\Delta\tan\beta)=
\Gamma^{R}_{H/A}\pm\Delta\Gamma^{R}_{H/A}$,
where $\Gamma_{H/A}=\frac{1}{2}(\Gamma_H+\Gamma_A)$.

For the method~(iii), $\tan\beta$ sensitivity is evaluated by solving
${\mathcal B}_h(\tan\beta\pm\Delta\tan\beta)={\mathcal B}_h^{\rm
obs}\pm\Delta{\mathcal B}_h^{\rm obs}$, where the accuracy of ${\mathcal
B}_h$ measurement is evaluated from the reference value by rescaling
the statistical factor by taking into account the change of the
expected number of events. 
The reference values for the $1\sigma$ accuracy of determining branching
ratios in the $b\bar{b}$ and $\tau^+\tau^-$ decay modes at the ILC with
$\sqrt{s}=250$~GeV and ${\mathcal L}=250$~fb$^{-1}$ are taken as 1.3\%
and 2\%, respectively, from the recent
reports~\cite{Asner:2013psa,Dawson:2013bba}\footnote{%
We note that an indication of $\sin(\beta-\alpha)\neq1$ in the THDM can
be obtained by measuring the absolute value of the $hVV$ couplings at
the ILC. 
At the ILC with $\sqrt{s}=250$~GeV and ${\mathcal L}=250$~fb$^{-1}$, the
best accuracy of the measurements can be expected for the $hZZ$ coupling
at 0.7\%~\cite{Asner:2013psa,Dawson:2013bba}.
}.

\begin{figure}[t]
 \centering
 \includegraphics[height=0.245\textwidth]{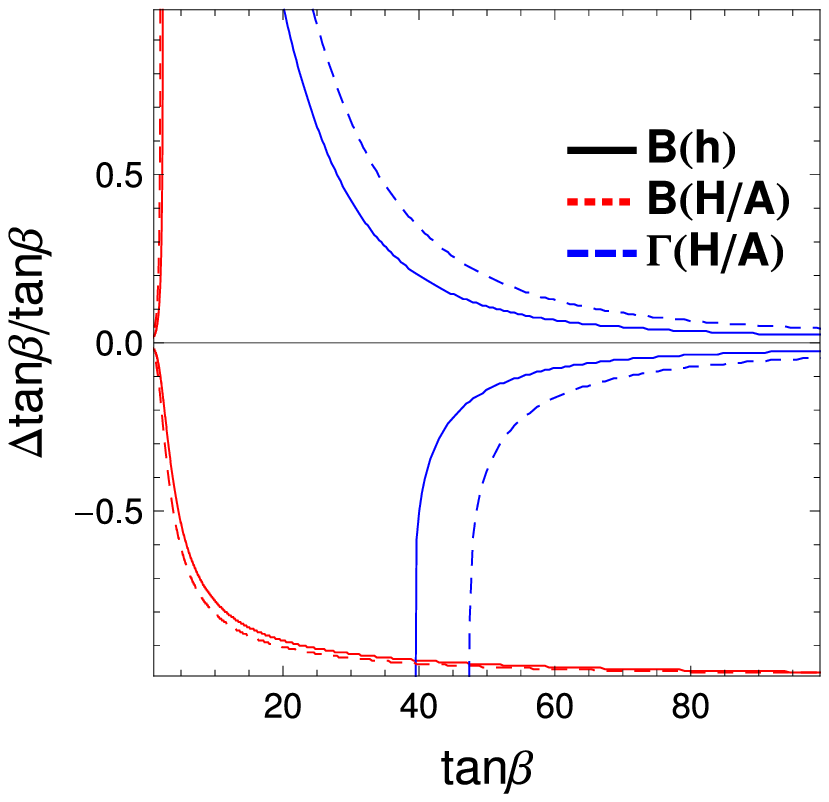} 
 \includegraphics[height=0.245\textwidth]{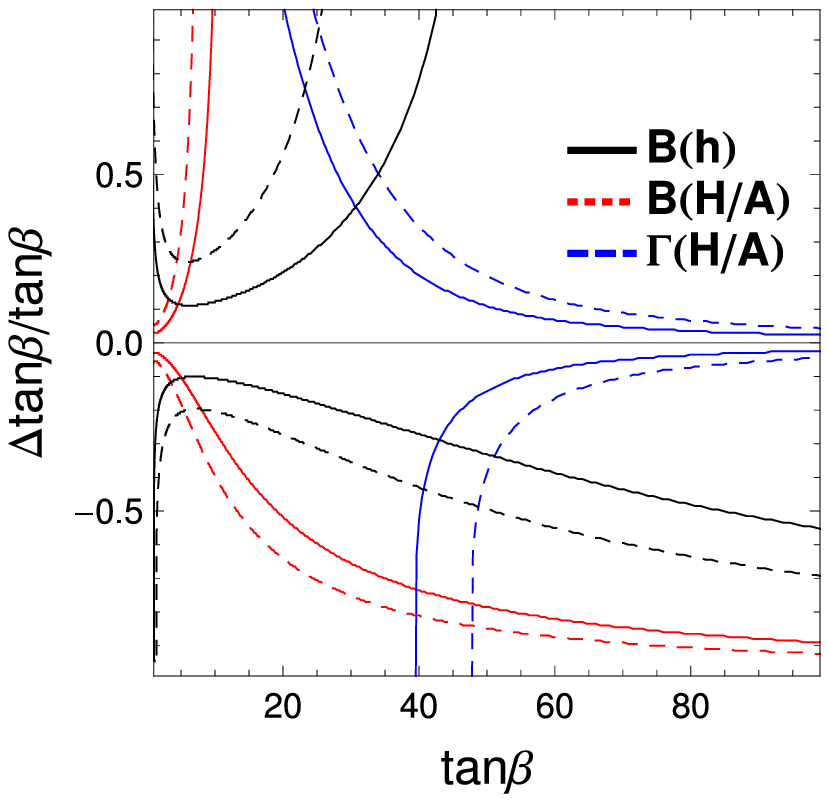} 
 \includegraphics[height=0.245\textwidth]{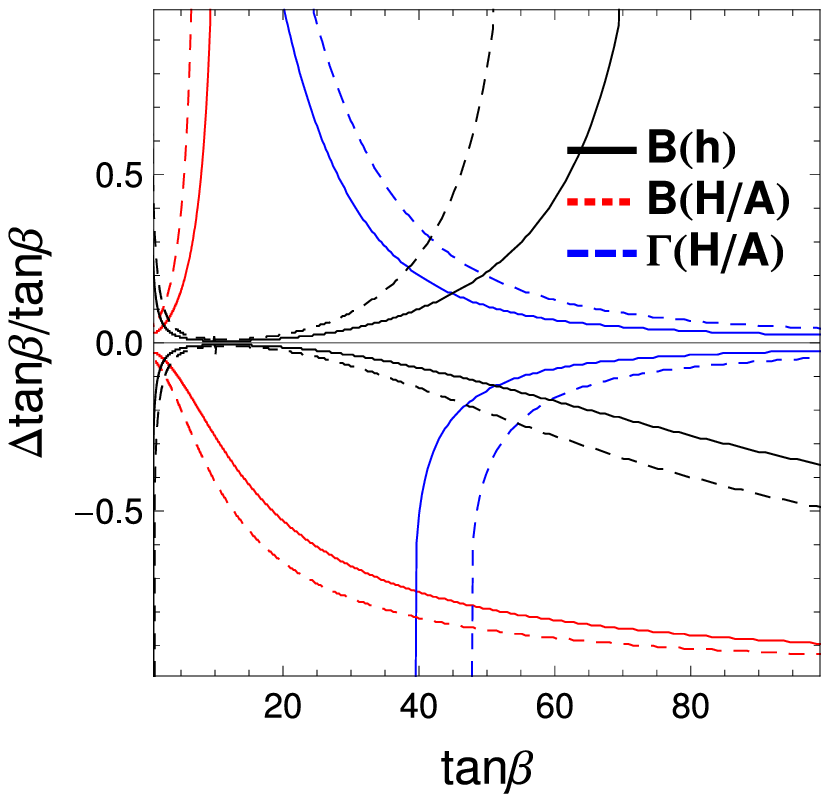} 
 \caption{Sensitivities to the $\tan\beta$ measurement in the Type-II
 THDM.
 From the left, $\sin(\beta-\alpha)=1$, $\sin^2(\beta-\alpha)=0.99$ with
 $\cos(\beta-\alpha)<0$, and $\sin^2(\beta-\alpha)=0.99$ with
 $\cos(\beta-\alpha)>0$ are taken, respectively.
}~\label{Fig:II}
\end{figure}

In Fig.~\ref{Fig:II}, our numerical results for the three methods are
shown for the Type-II THDM.
Left panel is for $\sin^2(\beta-\alpha)=1$, middle panel is for
$\sin^2(\beta-\alpha)=0.99$ with $\cos(\beta-\alpha)<0$, and right panel
is for $\sin^2(\beta-\alpha)=0.99$ with $\cos(\beta-\alpha)>0$.
$1\sigma$~(solid) and also $2\sigma$~(dashed) sensitivities are drawn as
a function of $\tan\beta$ for each method.
In the left panel, the method~(iii) does not work, since there is no
$\tan\beta$ sensitivity in the SM-like limit.
The method~(i) has good sensitivity in smaller $\tan\beta$ regions,
since there exists $\tan\beta$ dependence in ${\mathcal B}_{H/A}$ only
for these regions. 
The method~(ii) has good sensitivity in larger $\tan\beta$ regions, where
the widths can be directly measured. 
In the middle and right panels, when $\sin^2(\beta-\alpha)<1$, the
method~(iii) works very well in wide regions in $\tan\beta$. 

\begin{figure}[t]
 \centering
 \includegraphics[height=0.245\textwidth]{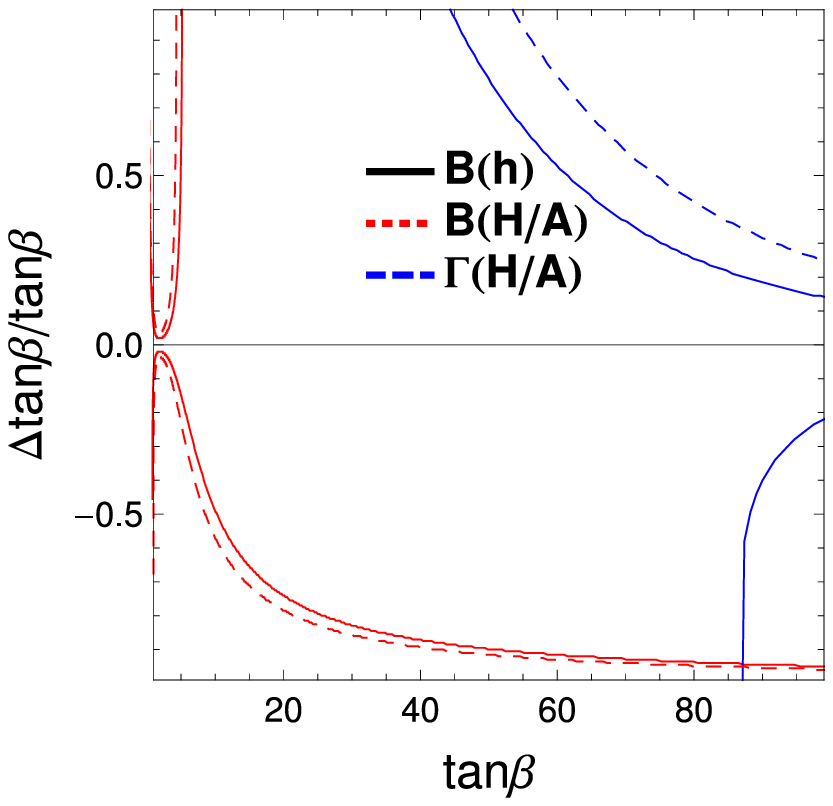} 
 \includegraphics[height=0.245\textwidth]{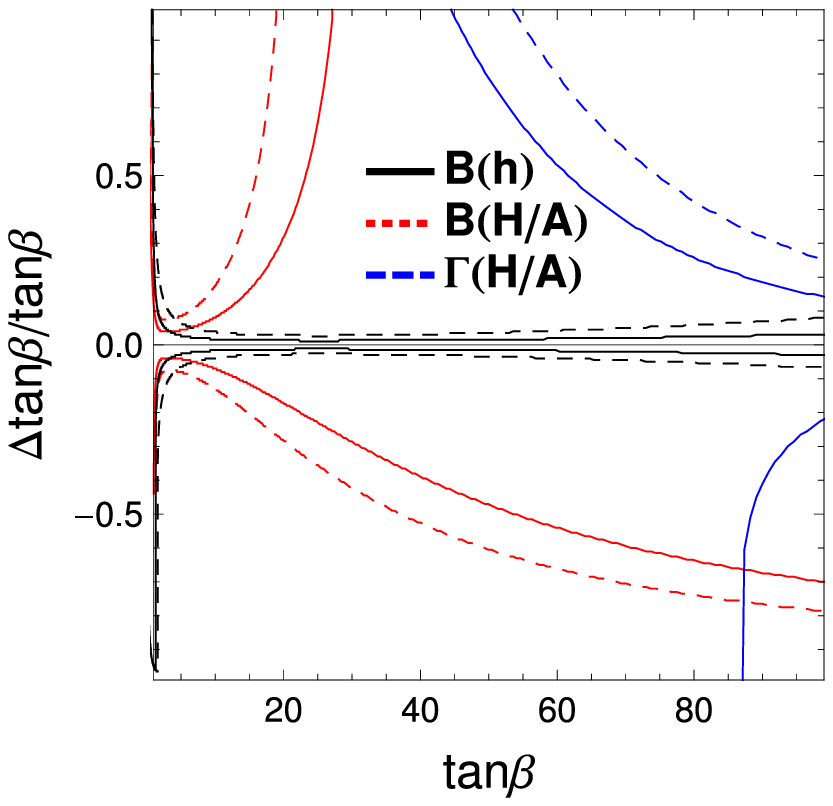} 
 \includegraphics[height=0.245\textwidth]{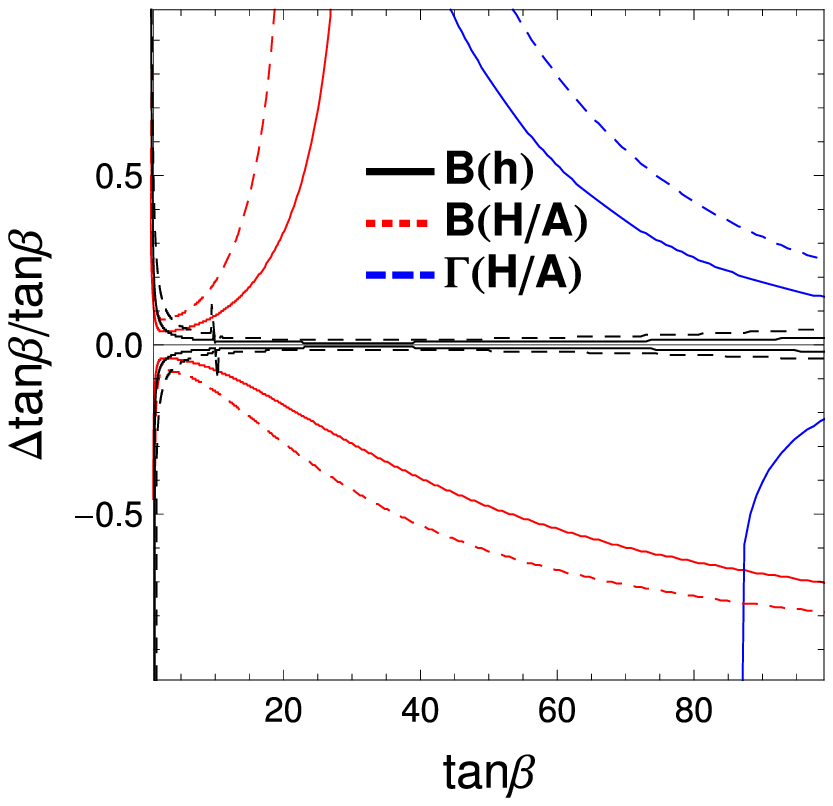} 
 \caption{The same as Fig.~\ref{Fig:II}, but for the Type-X
 THDM.}~\label{Fig:X}
\end{figure}

In Fig.~\ref{Fig:X}, numerical results for the three methods are
also shown for the Type-X THDM in the same manner as Fig.~\ref{Fig:II}.
The features of the three methods are similar to those for Type-II.

%%%%%%%%%%%%%%%%%
\section{Summary}

We have studied the sensitivities of $\tan\beta$ measurement by using
the complementary three methods at the ILC; (i) the branching ratio of
$H$ and $A$, (ii) the total decay width of $H$ and $A$, and (iii) the
branching ratio of $h$. 
The first two methods utilize the direct observation of the additional
Higgs bosons, $H$ and $A$. 
Therefore, these methods are available if the production process of
$e^+e^-\to HA$ is kinematically accessible. 
The last method utilizes the precision measurement of the branching
ratio of $h$ at the ILC. 
Although the method is available only for the case with
$\sin(\beta-\alpha)<1$ where $\tan\beta$ dependence can be seen in the
branching ratio of $h$, the method has better sensitivity for
determining $\tan\beta$ than the other methods in a wide range of
$\tan\beta$.

\bigskip % extra skip inserted
%%%%%%%%%%%%%%%%%%%%%%%%%%%%%%%%%%
\begin{acknowledgments}
The author would like to thank Shinya Kanemura, Koji Tsumura, and also
 Kei Yagyu for fruitful discussions and collaborations.
The work was supported in part by Grant-in-Aid for Scientific Research,
 No.\ 24340036 and the Sasakawa Scientific Research Grant from The Japan
 Science Society.
\end{acknowledgments}

\bigskip % extra skip inserted

\end{document}